# Survey Insights on M365 Copilot Adoption

Muneera Bano[1], Didar Zowghi[1], Jon Whittle[1], Liming Zhu[1], Andrew Reeson[1], Rob Martin[2], Jen Parsons[3]

[1] CSIRO's Data61, Australia
[2] IM&T, CSIRO, Australia
[3] Digital Office, CSIRO, Australia

firstname.lastname@csiro.au

## Abstract

Australia's National Science Agency conducted a six-month trial of M365 Copilot starting in January 2024 as part of an Australian Government initiative. 300 licenses were distributed across CSIRO using a persona-based approach to ensure diversity of roles and attributes among participants. As a scientific research organisation with a unique operational context, our objective was to study the use of M365 Copilot on enhancing productivity, efficiency, and creativity as well as overall satisfaction and ethical considerations. This paper presents the results from two surveys, conducted before and after the trial. The results showed mixed outcomes, with participants reporting improvements in productivity and efficiency in structured tasks such as meeting summaries and email drafting. Gaps were identified in advanced functionalities and integration, which limited Copilot's overall effectiveness. Satisfaction levels were generally positive, but user experiences varied in certain areas. Ethical concerns, particularly around data privacy became more pronounced post-trial. These findings highlight Copilot's potential while highlighting the need for further refinement to meet users' diverse needs.

## 1. Introduction

M365 Copilot is introduced as an everyday artificial intelligence (AI) companion designed to enhance productivity and creativity across various applications (Kytö 2024). It integrates AI capabilities into Microsoft's core products like GitHub, Microsoft 365, Bing, and Edge, aiming to streamline coding, transform workplace productivity, redefine search experiences, and provide contextual assistance .(Chen 2024; Grover 2024; Kytö 2024) The Copilot experience, which combines web context, work data, and real-time PC activities, prioritizes privacy and security. It's integrated into Windows 11, Microsoft 365, and Edge, offering seamless accessibility and evolving capabilities. The roll-out includes new features in Windows 11 and Microsoft 365, emphasizing enhanced AI experiences in common applications and introducing new Surface devices optimized for these AI tools.[1]

The Australian Government collaborated with Microsoft to explore generative AI in public services[2]. A six-month trial of M365 Copilot was planned to run from January to June 2024, involving Australian Public Service (APS) staff to enhance productivity. Over 50 APS agencies including Commonwealth Scientific and Industrial Research Organisation (CSIRO)[3] participated in a six-month trial of M365 Copilot[4]. The trial aimed to enhance productivity, develop AI skills, and explore safe, ethical use cases for generative AI in government operations. Over 7,400 public servants were involved, using AI for tasks like meeting summaries and internal correspondence.

---

[1] https://blogs.microsoft.com/blog/2023/09/21/announcing-microsoft-copilot-your-everyday-ai-companion/
[2] https://www.pm.gov.au/media/australian-government-collaboration-microsoft-artificial-intelligence
[3] https://www.csiro.au/en/
[4] https://www.dta.gov.au/blogs/aps-trials-generative-ai-explore-safe-and-responsible-use-cases-government

At CSIRO, alongside the six-month trial to assess the impact of M365 Copilot within the organisation, a parallel research study was also conducted. The research objective for conducting a mixed-method scientific study on the M365 Copilot trial was to study and explore its impact on productivity, efficiency, effectiveness, and ethical concerns within CSIRO. This approach was essential to capture both quantitative and qualitative data, providing a holistic understanding of user experiences and contextual nuances. We conducted two surveys with mixed-method study design. Pre-trial surveys were conducted with 244 participants to gauge initial expectations and perceptions of Copilot, followed by post-trial surveys involving 177 participants to assess changes in perceptions and experiences after using the tool.

The results revealed both opportunities and challenges in integrating the tool into professional workflows. While participants initially had high expectations for improvements in productivity and efficiency, post-trial feedback was mixed. Copilot showed promise in enhancing structured tasks, such as meeting summaries and email drafting. Post trial, ethical concerns, prominently data privacy, emerged as significant factor impacting user trust. Overall, while M365 Copilot demonstrated positive potential, addressing these challenges will be essential to maximise its impact in professional environments.

This research provides valuable contributions through a comprehensive, real-world evaluation ofM365 Copilots in the context of a scientific research organisation, specifically through pre- and post-trial surveys. By capturing user perceptions before and after the trial, the surveys offer critical insights into how M365 Copilots affect user perception of productivity, efficiency, and effectiveness, while also addressing shifts in user satisfaction and ethical considerations. These findings deepen our understanding of AI adoption in scientific organisation and offer practical guidance for researchers, practitioners, and policymakers on the responsible implementation of AI technologies.

This paper is structured as following: Section 2 presents the background and a review of existing literature on AI Copilots, exploring their potential applications, benefits, and challenges across various domains. Section 3 outlines the research methodology, detailing the design and execution of pre-trial and post-trial surveys used to gather quantitative and qualitative insights from participants. Section 4 focuses on the results, providing a thorough analysis of the data collected from both surveys. Section 5 discusses and compare results from both surveys. Section 6 addresses threats to validity. Finally, Section 7 concludes the paper, summarising the findings and offering directions for future research.

## 2. Background

The integration of AI into professional settings has been anticipated to alter workflows, with AI Copilots emerging as a notable development (Coyle and Jeske 2023; Sellen and Horvitz 2024). AI Copilots, Microsoft 365 Copilot[5], Google's Gemini[6] and Anthropic's Claud 2[7], are designed to assist users by providing real-time suggestions, automating routine tasks, and supporting decision-making processes. These tools utilise advanced machine learning algorithms to interpret user inputs and generate responses that align with the given context, which may contribute to increased efficiency and productivity. Typically powered by large language models (LLMs) like OpenAI's Codex and GPT-4, AI Copilots are being deployed in domains such as software development, marketing analytics, and knowledge work, where they are intended to complement human capabilities by managing routine tasks and streamlining processes (Peng et al. 2023; Coyle and Jeske 2023; Bano et al. 2024).

Research has reported varying outcomes regarding the potential of AI Copilots to enhance productivity across different fields. In software development, GitHub Copilot has been associated with efficiency improvements, with some studies indicating time savings in completing coding tasks (Peng et al. 2023). In marketing analytics, AI Copilots have been used to assist with data processing and decision-making, potentially reducing the time required for tasks like campaign optimisation (Coyle and Jeske 2023).

---

[5] https://blogs.microsoft.com/blog/2023/09/21/announcing-microsoft-Copilot-your-everyday-ai-companion/
[6] https://blog.google/technology/ai/google-gemini-ai/
[7] https://www.anthropic.com/news/claude-2

Additionally, knowledge workers, such as consultants and writers, have reported using AI Copilots to automate routine tasks, potentially allowing more focus on higher-level work (Dell'Acqua et al. 2023) (Ziegler et al. 2024). Research has demonstrated that large language models , such as ChatGPT 3.5 and GPT-4, have a promising impact on qualitative research by enhancing human capabilities rather than replacing them, as their collaboration with human analysts leads to improved outcomes through complementary strengths in classification and reasoning (Bano et al. 2024; Bano, Zowghi, and Whittle 2023). However, these tools also raise several concerns, including data privacy, intellectual property, and the risk of over-reliance on AI-generated content, all of which warrant further examination (Pearce et al. 2022; Moroz, Grizkevich, and Novozhilov 2022).

The existing literature on AI Copilots provides a nuanced view of their applications across various professional domains, highlighting both their strengths and limitations. In programming and software development, GitHub Copilot has been evaluated for its ability to assist developers by generating code suggestions and solving algorithmic problems. While it is shown to enhance productivity, particularly for expert developers, it also frequently produces buggy code and raises concerns about code quality, security, and ownership. Studies suggest that GitHub Copilot requires responsible use and thorough testing to mitigate these issues, with security risks being a particular area of focus (Dakhel et al. 2023) (Grover 2024; Horne 2023).

In other domains such as healthcare and education, AI Copilots like GPT-4 and M365 Copilot have demonstrated potential benefits, though with varying outcomes. In healthcare, ChatGPT-4 has shown strong performance in simplifying medical jargon and answering clinical questions, but its vulnerability to adversarial prompts highlights the need for ongoing security improvements (Tepe and Emekli 2024; Hannon et al. 2024; Semeraro et al. 2024). In education, these tools have been praised for improving student engagement and providing tailored learning experiences, although concerns about reliability and the opaque decision-making processes (black-box nature) persist (Hassani and Silva 2024; Chen 2024). Across all domains, the literature calls for continuous development and enhanced safeguards to address the ethical and security implications associated with the deployment of AI Copilots, particularly in sensitive fields like healthcare and cybersecurity (Hannon et al. 2024; Menz et al. 2024).

In research literature, productivity, efficiency, and effectiveness are interrelated concepts often used to evaluate performance across domains. Productivity typically refers to the quantity of outputs produced relative to inputs, such as tasks completed or publications generated, and is commonly used to assess organisational or technological contributions (Pritchard 1995). Efficiency, on the other hand, focuses on optimising resource utilisation, time, effort, or cost to achieve desired outputs with minimal waste, a critical factor in domains such as healthcare and software development (Lovell 1993). Effectiveness emphasises the quality and impact of outcomes, assessing whether objectives are met, as seen in AI studies evaluating decision-making accuracy or clinical support (Neely, Gregory, and Platts 1995). While these concepts are distinct, their overlap and domain-specific interpretations often create inconsistencies in their application, particularly in evaluating AI systems like Copilots, which must balance high output, resource optimisation, and goal alignment.

AI Copilots can increase both productivity and efficiency by enabling users to complete tasks faster and more accurately, reducing the need for manual labour, and allowing individuals to focus on tasks that require human creativity and decision-making (Ziegler et al. 2024; Sun, Che, and Wang 2024). In the context of AI Copilots, productivity is inherently shaped by the specific demands and priorities of each domain, with its definition varying based on the nature of tasks and desired outcomes. In software development, productivity often centres on the efficiency and accuracy of coding, measured through metrics such as lines of code written, debugging time, and feature implementation. AI Copilots like GitHub Copilot significantly enhance productivity by automating routine coding tasks, enabling developers to focus on complex problem-solving and higher-level design (Bird et al. 2022; Dakhel et al. 2023). In marketing, productivity is defined by the capacity to optimise campaigns, generate creative content, and derive actionable insights from data, with tools like Google Bard and Microsoft Copilot streamlining these processes and improving engagement metrics (Coyle and Jeske 2023). For knowledge work and research, productivity emphasises the synthesis of information, report generation, and decision-making efficiency, where tools like ChatGPT excel by reducing cognitive load, automating repetitive tasks, and providing insights that support innovation (Sellen and Horvitz 2024). These domain-specific applications of AI

Copilots demonstrate their adaptability and effectiveness while also emphasising the need for tailored metrics to evaluate their impact accurately within diverse professional contexts.

While the productivity benefits of AI Copilots are theoretically clear, they also introduce a range of risks and ethical challenges. Studies have shown that AI Copilots can inadvertently generate insecure or incorrect solutions, posing risks in domains where accuracy is critical. For instance, in (Pearce et al. 2022) the authors found that approximately 40% of the code generated by GitHub Copilot contained security vulnerabilities, raising concerns about the tool's reliability in sensitive software development environments. Moreover, AI Copilots also raise ethical issues related to privacy, data governance, and intellectual property. As these tools often rely on vast amounts of training data, which can include proprietary or sensitive information, their widespread use poses risks regarding data security and privacy (Coyle and Jeske 2023). Additionally, the intellectual ownership of AI-generated content is still a grey area, with concerns over whether AI Copilots may inadvertently infringe on copyrighted material (Moroz, Grizkevich, and Novozhilov 2022).

Another ethical concern is the potential for de-skilling and automation bias (Nazareno and Schiff 2021). As AI Copilots handle more routine tasks, there is a risk that human users may become overly reliant on these systems, leading to reduced vigilance and a decline in critical skills. This is particularly concerning in industries like software development, where maintaining a deep understanding of coding principles is crucial for quality control and innovation (Sellen and Horvitz 2024). Furthermore, over-reliance on AI-generated outputs can lead to automation bias, where users uncritically accept suggestions without sufficient scrutiny, potentially exacerbating errors or biases inherent in the AI models (Horne 2023).

Despite the rapid development and integration of AI Copilots in many domains, there is a notable gap in studies examining their potential to enhance productivity and innovation within scientific research. Relatively few studies have investigated AI Copilots' potential to accelerate scientific discovery, streamline research processes, or assist researchers in managing complex data and analysis workflows (Bano et al. 2024; Bano et al. 2023; Bano, Zowghi, and Whittle 2023). AI Copilots hold immense potential to transform scientific research, though domain-specific evidence supporting their integration remains limited (Bano et al. 2024). Australia's strategic initiatives highlight AI's ability to drive innovation across industries such as healthcare, agriculture, and manufacturing, emphasising principles of data privacy, accountability, and transparency (Hajkowicz et al. 2023). However, explicit discussions about AI Copilots or generative AI systems in the context of scientific research remain rare. One recent study by Toner-Rodgers (Toner-Rodgers 2024), provide critical insights into the impact of AI in research contexts. Their findings demonstrate that AI-assisted scientists achieve substantial productivity gains, including a 44% increase in material discoveries, a 39% rise in patent filings, and a 17% boost in product prototypes. However, disparities persist, as lower-performing researchers benefit less and express concerns about reduced creativity and job satisfaction, underscoring the need for AI tools to complement domain expertise effectively (Bano, Zowghi, and Whittle 2023). This lack of research presents an important opportunity for the scientific community to examine how AI Copilots might specifically impact research productivity, efficiency, and the quality of scientific outputs.

The novelty of our research lies in its duration and organisational context. Our study consists of two surveys: pre-trial and post-trial, conducted six months apart, offering both quantitative and qualitative insights into the adoption of M365 Copilot in an organisational context focused on scientific research. This extended trial period allows us to capture evolving perceptions and experiences. Moreover, the organisational context of scientific research adds an important dimension to our study, as it involves high levels of creativity and complex problem-solving. Understanding whether AI Copilots can enhance productivity, efficiency, and effectiveness, while maintaining user satisfaction and addressing ethical concerns, is particularly important in this field. The findings from this unique setup aim to inform researchers, practitioners, government bodies, and other organisations on the responsible and effective integration of AI technologies, providing a more holistic view of their potential and challenges in real-world environments.

## 3. Research Methodology

We utilised a mixed-methods approach, planned over a six-month trial of M365 Copilot within CSIRO to provide broader insights into participants' experiences. All user-reported data was anonymised to comply with CSIRO's human ethics research standards, ensuring participants could provide feedback freely, without any pressure or concern regarding their trial participation or license access. The study focused on user-perceived productivity, efficiency, and effectiveness, allowing participants to define these concepts based on their understanding within their professional context. CSIRO's Information Management and Technology (IM&T) division, and Digital Office, were mainly responsible for running the M365 Copilot trial by managing technical details related to licensing and ensuring seamless deployment across the organisation. These two units provided the support for data collection, ensuring that the research team had the necessary infrastructure and technical assistance.

The 300 participants for the trial (which were the target research participants for this study) were selected through a systematic persona-based approach, which was a significant methodological strength, ensuring the study captured a diverse range of experiences and needs. This diversity enriched the feedback and enhanced the validity and applicability of the study's findings, guiding the effective and inclusive implementation of M365 Copilot within the organisation. Participants were selected based on personas reflecting different job functions, seniority levels, and diversity characteristics. This approach ensured representation across various roles, such as senior leaders, legal advisors, researchers, IT managers, and operational staff, addressing specific needs and challenges unique to each role.

### 3.1 Pre-Trial Survey

The pre-trial survey was designed to gather baseline data on user expectations, existing workflows, and initial attitudes towards M365 Copilot. This structured questionnaire was administered to all participants at the outset of the trial, with 244 responses received. The survey focused on participants' anticipated productivity and efficiency gains, along with their initial thoughts on potential ethical concerns related to the use of AI Copilots. This baseline data set a reference point for understanding how Copilot could impact their work processes.

### 3.2 Post-Trial Survey

The post-trial survey aimed to assess changes in user perceptions after six months of using M365 Copilot. It was administered to all trial participants at the end of the trial, with 177 responses received, mirroring the pre-trial survey to enable a comparative analysis of expectations versus actual outcomes. This survey included additional questions on user satisfaction, ethical considerations, and any challenges encountered during the trial, providing a comprehensive overview of Copilot's impact on productivity, efficiency, and effectiveness over time.

### 3.3 Data Collection and Analysis

The research team designed the questionnaires for the surveys, in collaboration with the IM&T/Digital Office team, who were responsible for digital form designs on CSIRO platforms, and participant data collection. For compliance with Human Ethics Research rules, the research team was not to have any direct contact with trial participants to maintain confidentiality. All collected data were de-identified by IM&T/Digital Office after each phase and then handed over to the research team for analysis. All communications and interactions with participants were managed through IM&T/Digital Office. Participation in the research was entirely voluntary, with assurances that it would not affect trial participation. All data collected were self-reported and obtained with consent, ensuring privacy and respect for the participants' work environment.

The lead researcher directed the data analysis for the pre-trial and post-trial surveys. This process involved systematically identifying themes and patterns within the responses to each question. The collected data were meticulously coded and synthesised to build a comprehensive analysis addressing the research objectives related to productivity, efficiency, and ethical concerns. To ensure the reliability of the findings, peer reviews were conducted by other senior team members, providing critical insights and validating the analysis and interpretations of data.

### 3.4 Demographics of the trial participants

The trial for M365 Copilot at CSIRO used a persona-based approach to distribute licenses to individuals who expressed interest in participating. This strategy ensured a diverse and representative sample, capturing a broad range of perspectives and use cases, thereby enhancing the validity and applicability of the study's findings. Table 1 provides a summary of the different personas, their associated job roles, and the specific needs that guided the distribution of M365 Copilot licenses during the trial.

*Table 1 – Personna Archetypes*

| Persona Title | Overview | Job Roles | Needs |
|---|---|---|---|
| **Senior Leaders** | Direct strategic portfolios, manage organisational change, and oversee performance and growth. | Executive Team Member, Director, Deputy Director | Data for informed decision-making, tools for team engagement, and effective collaboration tools. |
| **Legal & Compliance Advisor/Manager** | Provide legal advice, manage legal risks, and oversee compliance. | Legal and Contracts Manager, Legal Counsel | Tools for tracking and auditing content changes, intelligent document analysis, enhanced data security. |
| **Researcher** | Conduct research, publish findings, and collaborate with industry. | Research Consultant, Research Scientist, Research Engineer | Advanced data management tools, support for drafting publications, collaboration tools for multidisciplinary teams. |
| **Research Manager** | Lead team operations, manage resource allocation, and oversee project outcomes. | Research Program Director, Group Leader, Team Leader | Tools for analysing project risks, tracking project status, and enhancing team collaboration. |
| **Research/Operations Support** | Deliver support across scientific and operational divisions, ensuring effective service provision. | Executive Officer, Administrative Officer | Meeting summarisation tools, task simplification for routine activities, effective collaboration tools. |
| **IT Advisor/Manager** | Manage IT strategy, oversee cybersecurity, and implement IT solutions. | IT Manager, IT Advisor, IT Engineer | Simplification of routine IT tasks, continuous learning tools, system event monitoring. |
| **Operational Services** | Manage internal and external communications, property operations, and policy assurance. | Enterprise Support Areas, Corporate Affairs, Governance | Data for informed decision-making, streamlined internal audits, efficient resource allocation. |
| **Finance Advisor/Manager** | Manage funding allocations, provide financial modelling, and ensure compliance. | Finance Manager, Finance Officer | Efficient financial forecasting, proactive monitoring of project funding, enhanced collaboration on financial insights. |
| **Business Development Manager** | Explore commercialisation opportunities, identify partnerships, and manage funding. | Business Development Manager, Commercial Manager | Partnership discovery tools, AI-assisted proposal writing, strategic planning support. |
| **People Advisor/Manager** | Manage recruitment, employee relations, and workforce planning. | Enterprise Executive Manager, HSE Manager | Advanced analytics for workforce planning, tools to streamline recruitment, insights for employee engagement strategies. |
| **Strategy & Analytics** | Conduct data analysis, performance management, and strategy planning. | Strategic Business Analyst, Data Analyst, Strategy & Performance Manager | Data for informed decision-making, enhanced data visualisation, AI-assisted strategy development. |

The trial participants represented a diverse demographic and professional profile, providing a comprehensive view of M365 Copilot's impact across different groups. The trial participants included a diverse range of personas (see Figure 1), with researchers, operational services, and IT advisors/managers being slightly overrepresented, reflecting the organisational emphasis on these roles. Smaller groups, such as senior leaders, legal and compliance advisors, and strategy and analytics roles, were proportionally represented based on their presence within the organisation. This distribution ensured a representative sample, providing valuable insights into Copilot's applicability across diverse professional contexts.

Gender distribution was relatively balanced, with a nearly equal number of participants identifying as women and men, while a smaller proportion identified as non-binary or preferred not to disclose their gender (see Figure 2). In terms of age (see Figure 3), the majority of participants were within the 25-34 and 35-44 age brackets, making up a significant portion of the trial population, which suggests that the trial attracted mid-career professionals likely to be key adopters of innovative tools.

Figure 4 shows the varying levels of prior knowledge and expertise with generative AI among participants. The largest group had average knowledge, followed by those with advanced understanding, while smaller proportions identified as experts or reported limited experience. A minimal number of participants were unfamiliar with generative AI, highlighting a generally moderate to high baseline familiarity across the cohort.

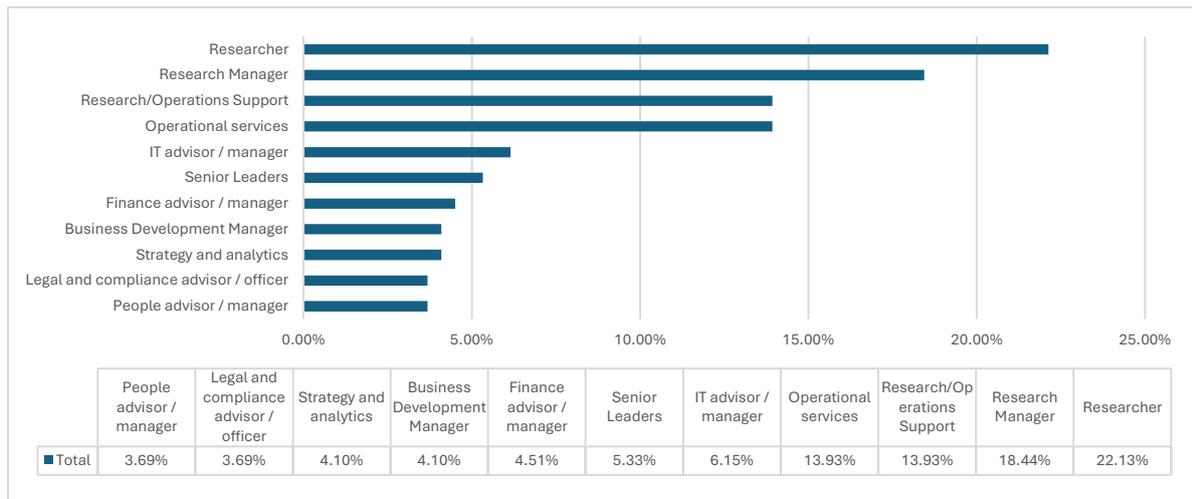

Figure 1. Percentages of the Persona for M365 license distribution

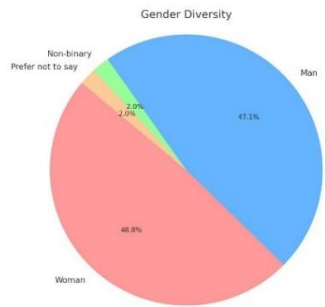

Figure 2. Gender diversity of trial participants

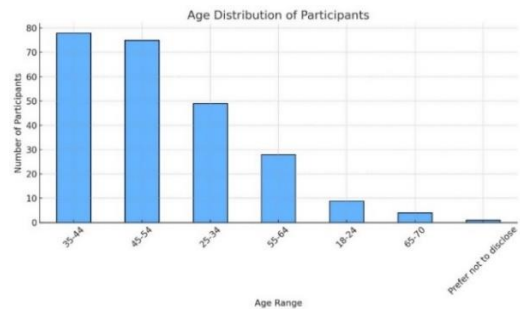

Figure 3. Age groups of trial participants

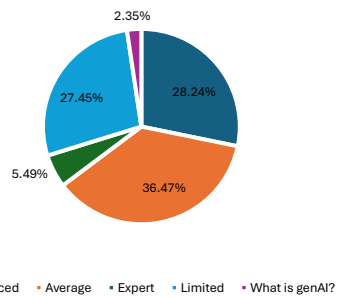

Figure 4. Level of prior knowledge and expertise with Generative AI

# 4. Results

## 4.1 Pre-Trial Survey

A total of 244 responses were received, providing valuable feedback on anticipated benefits such as increased efficiency, productivity, creativity, and time savings. Appendix A provides the questionnaire of Pre-Trial Survey.

### 4.1.1 Expectations from Copilot on Work aspects

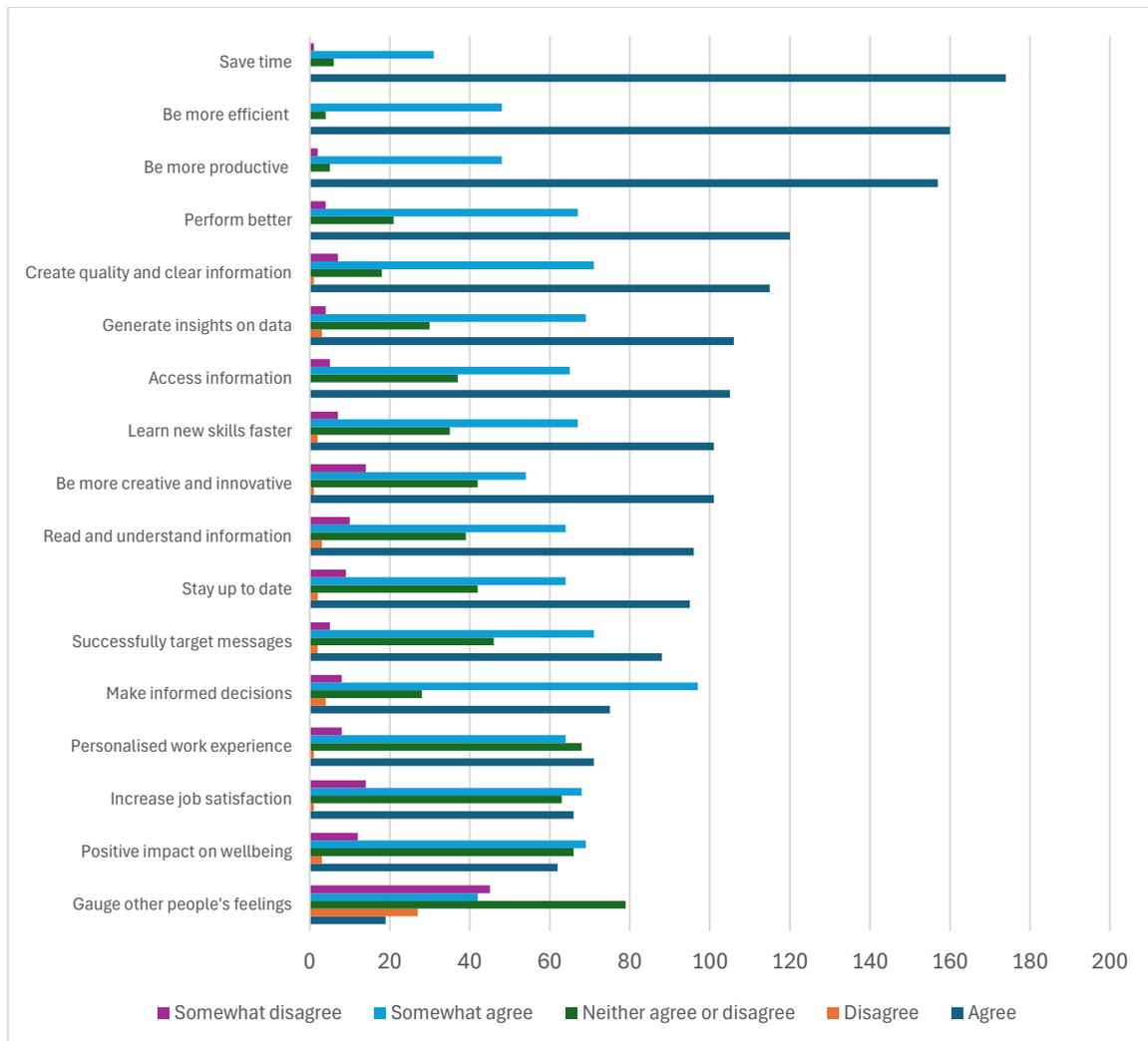

*Figure 5. Pre-trial expectations from M365 Copilot*

As shown in Figure 5, participants had high expectations for M365 Copilot across several dimensions. The most anticipated benefit was timesaving, with 210 participants (86%) strongly agreeing that it would reduce time spent on repetitive tasks. Efficiency was another key expectation, with 200 participants (82%) believing Copilot would enhance their ability to perform daily tasks more effectively. Similarly, 195 participants (80%) expected Copilot to improve access to information, aiding decision-making processes. Productivity expectations were also significant, with 190 participants (78%) anticipating that Copilot would streamline workflows and boost overall output. Additionally, 185 participants (76%) believed Copilot would improve job satisfaction by supporting workload management and performance. Finally, 180 participants (74%) expected Copilot to enhance creativity, enabling innovative approaches to problems and tasks. These numbers demonstrate the strong optimism surrounding Copilot's potential impact.

### 4.1.2 Use of Microsoft Tools Usage

Outlook and Teams are the dominant tools used a lot of the time compared to other products. The frequent use of Outlook and Teams aligns with participants valuing Copilot's integration in these tools, particularly for tasks like drafting emails and summarising meetings. This suggests that Copilot's functionality is most effective and impactful when integrated into tools that are already central to users' workflows.

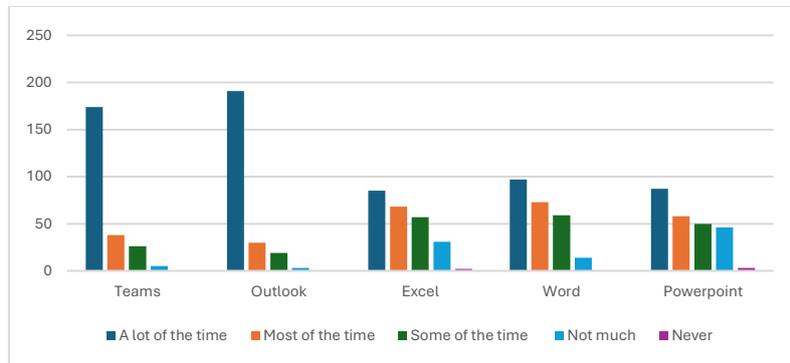

*Figure 6. Use of Microsoft tools*

### 4.1.3 Ethical and Integration Concerns

Most participants (220 participants, 90.2%) did not express significant ethical concerns about using Copilot. However, some participants (24 participants, 9.8%) raised questions about data privacy and the ethical implications of AI in decision-making. Concerns about integrating Copilot with existing tools and workflows were minimal, with 210 participants (86.1%) expressing no significant concerns. However, some participants (34 participants, 13.9%) were curious about its compatibility with Apple Mac OS and other less commonly used or customised system configurations.

### 4.1.4 Insights from Pre-Trial Survey

The pre-trial survey revealed a positive outlook among participants towards adopting M365 Copilot, with high expectations for improvements in efficiency, productivity, creativity, and job satisfaction. The demographic analysis showed a balanced representation across age groups and genders, with various roles within the organisation. While the majority did not express significant ethical or integration concerns, a few highlighted the need for attention to data privacy and compatibility with different operating systems.

The analysis of gender and expertise levels across various Microsoft products reveals notable trends. Women tend to have higher percentages of advanced and average expertise levels in most products, indicating proficiency. Men, however, have a higher percentage identifying as experts, especially in Excel and PowerPoint, suggesting a higher confidence or specialisation in these tools. Non-binary participants and those who prefer not to disclose their gender have lower representation across all expertise levels. These insights highlight a gender disparity in perceived or actual expertise levels, suggesting the need for targeted training and confidence-building initiatives to ensure balanced proficiency and representation across all gender groups.

## 4.2 Post-Trial Survey

171 participants responded to post-trial survey, covering multiple dimensions of their experience with Copilot. The responses captured detailed feedback on specific statements, time spent on various activities, and participants' concerns and perceptions about Copilot. Participants were asked to indicate their agreement or disagreement with statements regarding Copilot's impact on various aspects of their work (see Figure 7 and 8).

**4.2.1 Copilot's impact on work**

In terms of **efficiency**, 37% of participants somewhat agreed that Copilot enhanced their efficiency, while 18% agreed and 11% disagreed. This suggests that while a significant portion of participants found Copilot beneficial for efficiency, the mixed responses indicate potential areas for improvement, such as addressing specific inefficiencies, enhancing usability, or providing better integration into existing workflows.

Regarding **productivity**, 37% somewhat agreed that Copilot improved their productivity, with 21% agreeing and 11% disagreeing. Similar to efficiency, these results indicate a generally positive impact, though not without some reservations. In terms of **creativity**, 30% somewhat agreed that Copilot helped in this area, while 18% agreed and 11% disagreed. This indicates a more mixed reception regarding Copilot's role in enhancing creativity.

For **timesaving**, the responses were balanced, with both 37% agreeing and somewhat agreeing that Copilot helped save time. Only 5% disagreed with this statement. In terms of **job satisfaction**, 30% of participants neither agreed nor disagreed that Copilot improved their job satisfaction, while 30% somewhat agreed and 18% agreed. However, 12% somewhat disagreed, and 9% disagreed, indicating a diverse range of experiences.

These findings highlight Copilot's perceived benefits in terms of saving time for a majority of users, though there remains a segment that did not experience these benefits.

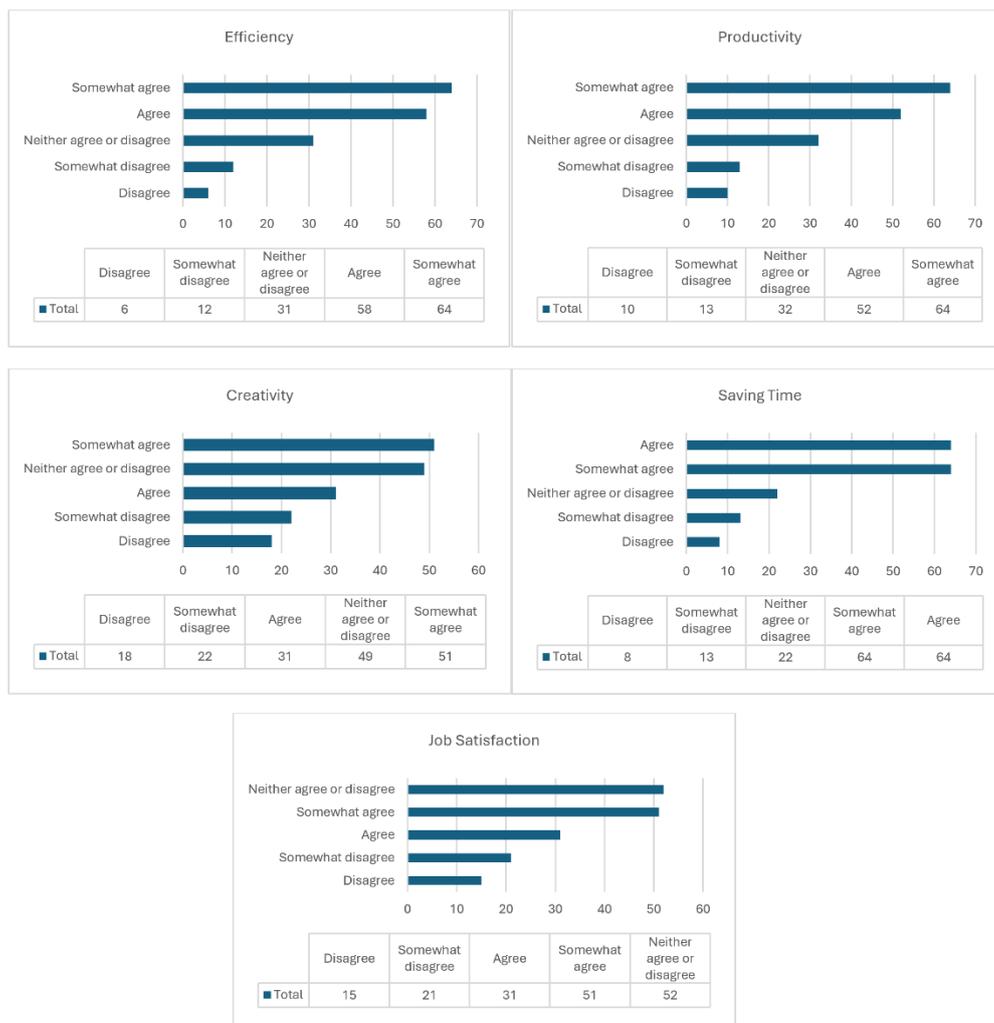

*Figure 7. Post-trial survey results for "efficiency", "productivity", creativity", "saving time" and "job satisfaction"*

### 4.2.2 Overall Satisfaction and Recommendations

The likelihood of recommending Copilot to colleagues or within the professional network averaged 7.13 out of 10, indicating a generally positive reception. The recommendation scores varied significantly, with a standard deviation of 2.44. This standard deviation highlights the spread of responses, showing that while many participants gave high scores, there were notable deviations, indicating variability in participants' perceptions of Copilot's effectiveness. Most recommendation scores were clustered between 6 and 10, reflecting a strong inclination toward recommending the tool. However, there were a few outliers below the score of 3, representing participants who were either dissatisfied or ambivalent about Copilot. Specifically, 7 participants provided scores of 0 or 1, representing 4.1% of the total 171 responses. While these outliers are relatively small in number and do not significantly impact the overall trend, they provide valuable insights into areas of concern that may need to be addressed. These participants generally disagreed that Copilot made them more efficient, productive, creative, or helped them save time. Their recommendation scores ranged from 0 to 1, indicating a strong reluctance to recommend Copilot.

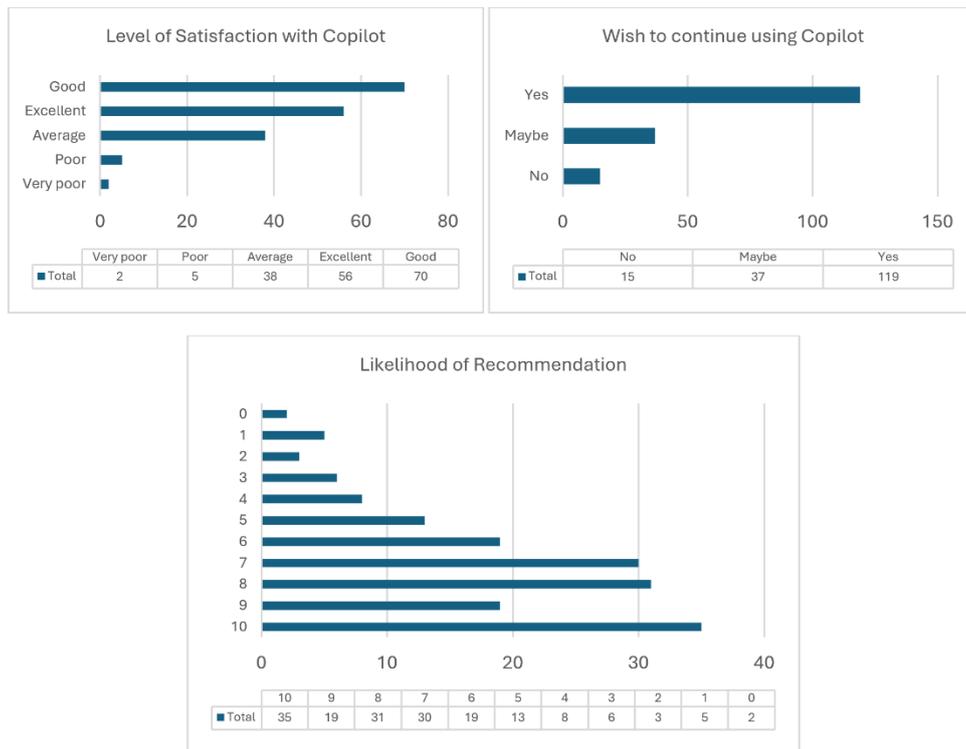

*Figure 8. Post-trial survey results for "level of satisfaction with Copilot", "continued use of Copilot", and "likelihood of recommendation"*

### 4.2.3 Copilot Impact on work activities

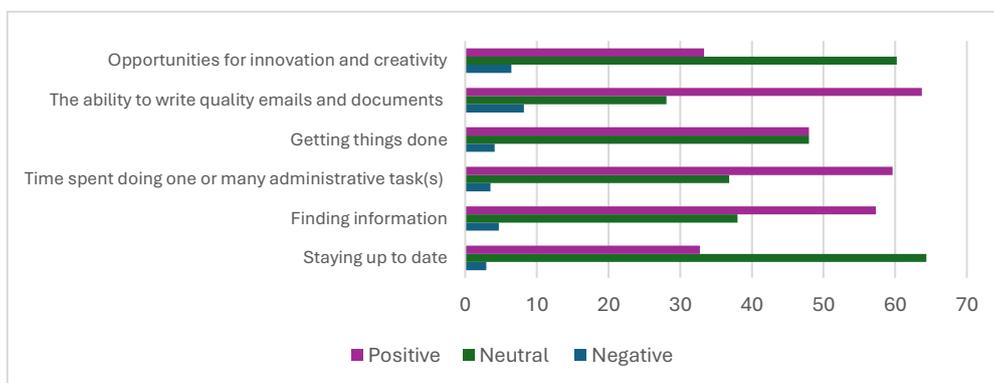

*Figure 9. Post-trial survey results for impact on work activities*

The graph in Figure 9 demonstrates a diverse range of perceptions regarding M365 Copilot's effectiveness, with clear areas of strength and opportunities for improvement. For "Opportunities for innovation and creativity," 33% of participants gave positive feedback, but a substantial 60% remained neutral, indicating a lack of a strong impact in this area. Negative responses were minimal at 7%, but the overwhelming neutrality suggests Copilot has yet to make a meaningful contribution to fostering creativity.

In contrast, "The ability to write quality emails and documents" emerged as a success, with 64% of participants reporting a positive impact, significantly outpacing the 28% neutral responses and only 8% negative feedback. This indicates that Copilot has effectively enhanced participants' efficiency in this area. Similarly, "Time spent doing one or many administrative tasks" also received a robust 60% positive feedback, with 36% neutral and only 4% negative, showcasing Copilot's capability in streamlining routine tasks.

"Getting things done" presents a more balanced outcome, with equal proportions of positive (48%) and neutral (48%) feedback, and just 4% negative responses. While the positive impact is notable, the lack of differentiation from neutral responses suggests room for improvement in demonstrating value in task completion. "Finding information" saw 58% positive responses, with 38% neutral and only 4% negative, reflecting moderate success but also potential to further enhance the tool's utility in this area.

"Staying up to date" received the weakest positive feedback, with only 32% of participants reporting a favourable impact, while a dominant 65% remained neutral, and 3% provided negative responses. This indicates a significant gap in Copilot's ability to provide value in this task, despite its overall effectiveness in other areas.

These percentages highlight Copilot's strong performance in tasks like email writing and administrative work, where positive feedback far outweighs neutral and negative responses. However, the neutral dominance in areas like staying up to date and fostering creativity reveals a critical need for targeted improvements to maximize Copilot's potential and better meet user expectations.

### 4.2.4 Ethical Concerns

Most participants (108, 63%) did not identify any ethical concerns related to using Copilot (see Figure 10). However, a small but notable number expressed concerns (31, 18%) or were unsure about potential ethical issues (29, 17%). Several ethical concerns were identified as captured in Table 2.

The analysis highlights key concerns and insights related to the use of Copilot, centred around themes of privacy, data security, inclusivity, its impact on jobs, and the need for improved guidance. Privacy concerns dominate, with participants raising issues about document accessibility without creators' knowledge, meeting recordings lacking clear storage protocols, and misconceptions about unauthorised access. Ethical concerns also emerged, particularly around plagiarism, attribution, and the need to address biases in AI training data to ensure transparency and prevent misleading outputs. Inclusivity was another critical theme, with Copilot struggling to handle multicultural contexts, leading to transcription inaccuracies. Additionally, participants expressed worries about its potential to justify administrative staff reductions, underscoring the need for ethical implementation. Finally, there is a call for clearer guidance to build trust in Copilot's corporate isolation features, demonstrating a broader need for user education and transparency in deployment.

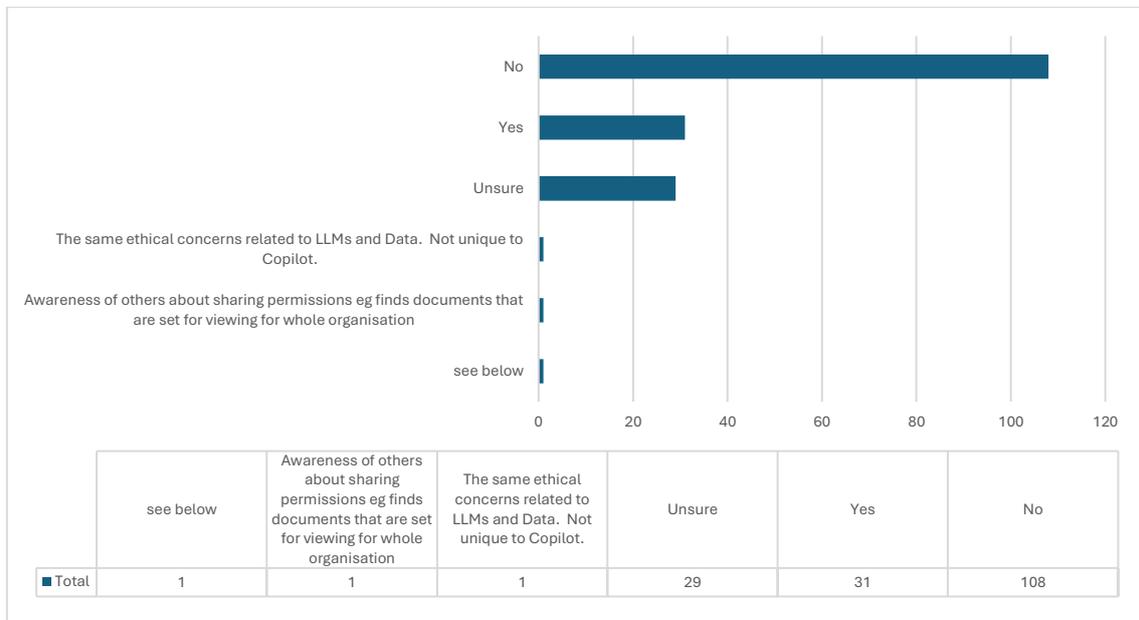

*Figure 10. Post-trial survey results for ethical concerns*

*Table 2 – Ethical concerns*

| Overarching Theme | Sub-Theme | Description |
|---|---|---|
| Privacy and Data Security | Document Accessibility and Awareness | Concerns about Copilot locating documents in CSIRO systems without creators' knowledge, raising questions about sensitivity and intended distribution. |
| | Meeting Recordings and Privacy | Emphasis on clear guidelines for storing and accessing Copilot-recorded meetings, including deletion protocols and differentiating official and candid records. |
| | Plagiarism and Attribution | Concerns about AI producing content used without proper attribution, raising ethical issues about originality and critical analysis of outputs. |
| | Unauthorised Access Misunderstanding | Misunderstanding of Copilot surfacing content users were authorised to view, seen as unauthorised access; need for clearer security protocol communication. |
| | Bias, Transparency, and Hallucination | Concerns about secure handling of sensitive data, addressing biases in AI training data, ensuring transparency, compliance, and preventing misleading outputs. |
| Impact on Jobs | Administrative Staff Reductions | Worries about Copilot being used to justify staff reductions, particularly in administrative roles. |
| Inclusivity | Anglo-Centric Bias and Multicultural Challenges | Difficulties with transcription accuracy in multicultural contexts and accents, raising concerns about inclusivity and accuracy. |
| Training and Guidance | Trust and Corporate Isolation | Need for clearer guidance and definitive answers to build trust in Copilot's corporate isolation capabilities. |

### 4.2.5 Statistical Analysis

Table 3 summarises the key findings from the statistical analyses conducted to explore the relationship between participants' perceptions of Copilot's productivity, their ethical concerns, and their likelihood of recommending the tool. It highlights insights into how agreement levels on productivity and the presence of ethical concerns influence recommendation behaviour. The results reveal that participants who perceive Copilot as productive are more likely to recommend it. Perceptions of productivity strongly influence recommendation behaviour, with differences observed across varying levels of agreement. Additionally, participants who raised ethical concerns about Copilot were less inclined to recommend it, highlighting the importance of addressing such issues to improve user acceptance.

Table. 3 - *Summary of Statistical Analyses Exploring the Relationship Between Perceptions of Productivity, Ethical Concerns, and Recommendation Likelihood for Copilot*

| Test | Purpose | Results | Interpretation |
|---|---|---|---|
| **Chi-Square Test** | To determine the association between agreement on Copilot's productivity and likelihood of recommending it. | Chi-square statistic: 15. p-value: 0.045 Degrees of freedom: 6 | Significant association exists between agreement on Copilot's productivity and recommendation likelihood (p < 0.05). |
| **ANOVA** | To compare average recommendation scores across levels of agreement on Copilot's productivity. | F-statistic: 4.57 p-value: 0.001 | Significant differences in recommendation scores across agreement levels suggest perceptions of productivity influence recommendation likelihood. |
| **T-Test** | To compare recommendation scores between participants with and without ethical concerns. | t-statistic: -2.94 p-value: 0.004 | Significant difference in recommendation scores, with participants who identified ethical concerns being less likely to recommend Copilot (p < 0.05). |

### 4.2.6 Insights from Post-Trial Survey

The post-trial survey results encompassing feedback from 171 participants, indicate a generally positive reception with significant variability in specific areas. Participants reported moderate to high levels of agreement on Copilot's impact on efficiency, productivity, creativity, and timesaving. The likelihood of recommending Copilot averaged 7.13 out of 10, with some notable outliers expressing dissatisfaction. Ethical concerns were identified by 18% of participants, primarily focused on document accessibility, privacy, uncritical use, and potential biases in the AI model. Statistical tests revealed associations between productivity perception and recommendation likelihood, as well as differences in satisfaction based on ethical concerns. While Copilot's positive impact on activities like staying up to date and getting things done is evident, areas such as writing quality emails and fostering innovation show room for improvement.

## 5. Pre-Trial and Post-Trial Surveys Results Comparison

In this section, we present insights into three critical dimensions of M365 Copilot's implementation: expectations, ethical concerns, and overall impact. These dimensions provide a comprehensive understanding of Copilot's performance and its reception among participants. Expectations reflect participants' initial optimism regarding Copilot's potential to enhance productivity, efficiency, and job satisfaction, and their comparison with post-trial outcomes reveals areas of alignment and gaps that need addressing. Ethical concerns surfaced as a significant theme during the trial, highlighting practical issues around privacy, data security, and fairness that are essential for fostering trust and ensuring equitable deployment. Overall impact, including productivity and timesaving, provides a holistic view of Copilot's tangible contributions to participants' workflows and satisfaction.

### 5.1 Level of Engagement and response Rate

The pre-trial survey included 244 participants, while the post-trial survey included 177 participants. The trial had 300 licenses distributed across CSIRO. The pre-trial survey achieved a high response rate indicating strong initial interest and engagement among participants. This high response rate suggests that the participants were keen to share their expectations and were motivated to contribute to the trial process from the outset.

The post-trial survey response rate dropped in comparison to the pre-trial survey. This reduction in participation may reflect several factors. Survey fatigue is a plausible explanation, particularly given the seven pulse surveys conducted by IM&T during the trial (though not part of this research), which may have contributed to participant disengagement. Additionally, some participants may have left the organisation during the trial, reducing the pool of potential respondents. Time constraints, technical difficulties, or a loss of interest due to their experiences—whether positive or negative—could have also played a role. These combined factors likely contributed to the lower response rate, despite concerted efforts to maintain engagement.

Participation in the research study was entirely voluntary, allowing participants to choose whether to provide feedback without any obligation or pressure. Considering that, despite the drop in the response rate of the post-trial survey, most of the participants remained engaged enough to provide feedback after the trial period.

### 5.2 Expectations and Initial Perception Evolution

Participants in the pre-trial survey had high expectations for improvements in efficiency and productivity, anticipating that M365 Copilot would significantly enhance their ability to perform daily tasks. They also expected boosts in creativity and considerable time savings, believing that Copilot would streamline processes and reduce the time spent on repetitive activities. Additionally, participants anticipated easier access to information, which they believed would aid in better decision-making. Overall, there was a strong expectation of improved job satisfaction due to better workload management and enhanced performance facilitated by Copilot.

In the post-trial survey, participants provided feedback on Copilot's impact, reflecting a generally positive reception in several key areas. Many participants acknowledged improvements in efficiency, noting that Copilot enhanced their ability to manage tasks and workflows. Productivity also saw positive feedback, with participants reporting that Copilot helped them perform better in their roles. However, feedback on creativity was more mixed, with some participants finding Copilot helpful in fostering innovation while others perceived less impact. Similarly, Copilot's contribution to timesaving was acknowledged, with participants highlighting its ability to streamline tasks and reduce time spent on repetitive activities. While Copilot met many expectations in areas like efficiency, productivity, and timesaving, its impact on creativity and broader job satisfaction was less consistent, suggesting areas for further refinement to align more closely with user needs and expectations. Before the trial, the majority of participants anticipated that Copilot would significantly enhance their job satisfaction by improving workload management and overall performance. However, post-trial feedback showed more mixed outcomes, with many participants expressing neutral or moderate agreement regarding its impact, and a smaller but notable group expressing dissatisfaction. While Copilot did meet many of the participants' expectations, particularly in terms of efficiency, productivity, and timesaving, the tool's impact on creativity and overall satisfaction was less consistent, highlighting areas for potential enhancement and better alignment with user expectations.

### 5.3 Misalignment of User Expectations vs. Actual Capabilities of M365 Copilot

User expectations for were high, with many anticipating significant enhancements in efficiency, productivity, and overall job satisfaction. They expected Copilot to streamline routine tasks, aid in complex decision-making, and seamlessly integrate with existing workflows. However, the actual capabilities of Copilot, while impressive, did not fully meet these expectations for all users. Some users reported notable improvements in specific areas like meeting summarisation and email drafting, yet broader productivity gains were often hindered by integration challenges and usability issues.

Many users did not explore the full potential of Copilot. Factors such as insufficient training, resistance to change, and a lack of awareness about advanced features may have led to underutilisation. Furthermore, the learning curve associated with mastering Copilot's capabilities added to the cognitive overload for staff already busy with their work tasks.

This gap between expectations and actual use highlights the need for comprehensive training and support, as well as strategies to manage cognitive load, to help users fully leverage the capabilities of AI tools.

### 5.4 Ethical Concerns

In the pre-trial survey, ethical concerns were minimal but did include issues related to data privacy and the ethical implications of AI in decision-making. Integration concerns were primarily focused on compatibility with non-standard setups such as Mac OS.

However, in the post-trial survey, ethical concerns became more pronounced. Participants highlighted issues such as document accessibility without the creator's knowledge, privacy concerns related to meeting recordings, and risks associated with the uncritical use of Copilot, including potential plagiarism. There were also significant concerns about unauthorised access to files, data security, and overall privacy. Additional worries included biases in AI training data, the potential negative impact on administrative staff, and the risk of AI generating inaccurate or misleading information.

These insights suggest that while initial concerns were relatively limited, the practical use of Copilot revealed deeper ethical and integration issues that need to be addressed to ensure secure and equitable deployment of the tool.

### 5.5 Participant Satisfaction and Recommendations

In the pre-trial survey, there was high optimism among participants regarding Copilot's potential to improve work efficiency, productivity, and job satisfaction. However, post-trial results showed that overall satisfaction was more measured, with the likelihood to recommend Copilot averaging 7.13 out of 10.

Despite this generally positive reception, there were notable outliers whose disagreement with Copilot's benefits in areas such as efficiency, productivity, creativity, and timesaving led to low recommendation scores ranging from 0 to 1. These outliers indicate that while many participants found value in Copilot, a notable minority experienced dissatisfaction, highlighting areas where Copilot's performance could be improved to better meet user expectations and needs.

### 5.6 Addition to the body of knowledge

The results of our study align with and extend the existing literature on the adoption of AI Copilots across various domains, offering fresh insights specific to the organisational context of scientific research. Prior research, such as (Coyle and Jeske 2023; Peng et al. 2023) has demonstrated the potential of AI Copilots to enhance productivity and efficiency, particularly in software development and marketing. Similarly, our participants reported notable improvements in efficiency and productivity post-trial, affirming these findings. However, our study also highlights mixed feedback on creativity and job satisfaction, echoing concerns raised by (Dell'Acqua et al. 2023; Ziegler et al. 2024) who observed that the benefits of AI Copilots often vary depending on the user's role and the specific demands of their work.

Ethical concerns, a recurring theme in prior literature, were also significant in our findings. Issues such as data privacy, transparency, and the potential for AI bias align with those discussed by (Moroz, Grizkevich, and Novozhilov 2022; Pearce et al. 2022). Our study expands on these insights by identifying specific ethical challenges encountered during the trial, such as document accessibility without creators' knowledge and unauthorised access perceptions.

Our research contributes new insights by examining Copilot's impact over a six-month trial in the unique context of scientific research. Unlike studies focused on domains such as education or software development, where task automation and decision-making support are prominent, our study explores how AI Copilots influence workflows requiring high levels of creativity and complex problem-solving. The findings suggest that while AI Copilots can effectively enhance productivity and streamline routine tasks, their integration in contexts demanding nuanced judgment and innovative thinking presents unique challenges. This underscores the need for tailored implementation strategies to maximise their potential and address context-specific barriers.

### 5.7 Contextual Insights on Productivity, Effectiveness, and Efficiency

The findings of our study underscore that productivity, effectiveness, and efficiency are inherently task- and tool-dependent, deeply contextual to the domain of work. In the organisational setting of scientific research, where workflows demand high levels of creativity, critical thinking, and complex problem-solving, the impact of AI Copilots like M365 Copilot varies significantly from more structured environments such as software development or marketing. For instance, while Copilot excelled at automating routine, structured tasks and improving document handling, its contributions in helping with creativity and innovative problem-solving were less pronounced, reflecting its alignment with predefined, procedural tasks rather than abstract, conceptual challenges.

This domain-specific dependency mirrors observations in existing literature, where productivity in software development is often evaluated through metrics like time saved in coding or debugging (Peng et al. 2023), while in marketing, it focuses on campaign optimisation and actionable insights (Coyle and Jeske 2023). Our study highlights that these measures are less applicable in scientific research, where productivity includes synthesising information, hypothesis generation, and exploratory work. The tools' ability to complement rather than replace human expertise becomes critical, as the balance between automation and cognitive engagement shapes perceived effectiveness and satisfaction.

Moreover, the effectiveness of AI Copilots is influenced not only by the nature of tasks but also by user engagement and tool utilisation. Participants who actively explored Copilot's advanced features reported greater perceived benefits, indicating that the tool's value is closely tied to user familiarity and application. Conversely, challenges such as insufficient training, resistance to change, and ethical concerns further contextualised its impact, as these factors often deterred participants from fully leveraging the tool. These insights reveal that while AI Copilots hold promise across domains, their integration must consider the unique demands and workflows of each context, with tailored strategies to maximise their relevance and utility.

## 6. Threats to Validity

### 6.1 Internal Validity

Internal validity in this study refers to the extent to which the observed effects can be attributed to the intervention, M365 Copilot, rather than external factors. Several potential threats to internal validity were identified. Selection bias may have influenced outcomes, as participants who volunteered for the trial could have pre-existing interests or biases toward AI technologies, shaping their experiences and feedback. Maturation effects are another consideration, as changes in productivity or attitudes over time could stem from natural progression rather than Copilot's direct impact. Additionally, the Hawthorne Effect may have played a role, with participants potentially altering their behaviour simply due to the awareness of being observed, which might inflate their perceptions of Copilot's effectiveness. Finally, repeated exposure to surveys, including pre-trial and post-trial surveys, in-depth interviews, and pulse surveys conducted by IM&T, could have sensitised participants to the study's objectives, influencing their responses and potentially skewing the results. These factors highlight the complexity of isolating Copilot's true impact from other influences.

### 6.2 External Validity

External validity in this study pertains to the generalisability of the findings to other settings, populations, or times. Several threats to external validity were identified. The sample representativeness poses a concern, as although 244 participants completed the pre-trial survey and 177 completed the post-trial survey, these samples—while relevant to CSIRO and similar organisations—may not fully represent the broader population of scientific researchers using AI tools. The study's context specificity is another factor, as it was conducted within CSIRO, whose distinct organisational culture and workflows may not directly translate to other research institutions. Additionally, the variability in participants' technological environments, including differences in their setups and familiarity with AI tools, may limit the applicability of the findings to settings with diverse technological infrastructures. These considerations highlight the need for caution when extending the study's conclusions to broader contexts.

### 6.3 Construct Validity

Construct validity in this study refers to how accurately the measures used reflect the intended constructs, such as efficiency, productivity, ethical considerations, and user satisfaction. Several threats to construct validity were identified. Operational definitions of these constructs might not have been uniformly understood by all participants, potentially leading to inconsistencies in their responses. The measurement instruments, primarily self-reported surveys, may not fully capture the complexity of participants' experiences with Copilot and are susceptible to biases like social desirability and recall bias. Additionally, the presence of confounding variables, such as concurrent organisational changes like new policies or additional training programs, could obscure the specific effects of Copilot, making it challenging to isolate its impact with precision. These factors highlight the importance of robust measurement tools and careful interpretation of the findings.

### 6.4 Addressing Threats to Validity

To address these validity threats, a mixed-method approach was employed, combining quantitative and qualitative data from pre-trial and post-trial surveys. We have separately conducted in-depth interviews following the completion of these surveys, and an experiment (not reported in this paper). This triangulation of data sources helped validate findings and provided a comprehensive understanding of Copilot's impact. However, awareness of these potential validity threats is essential for interpreting the results and guiding future research and implementation efforts.

## 7  Conclusion and Future Research Directions

The pre-trial survey revealed high expectations for M365 Copilot, with participants anticipating significant improvements in efficiency, productivity, and job satisfaction. Many were optimistic about Copilot's potential to streamline routine tasks, assist with document and meeting management, and enhance creativity, with minimal ethical concerns focused primarily on data privacy and AI's role in decision-making. However, the post-trial survey presented a more tempered perspective, indicating that while some expectations were met, there were notable gaps. While participants appreciated features like meeting summaries and email drafting, unmet expectations regarding advanced functionalities and integration issues were common. Ethical concerns also grew, particularly around unauthorised access to sensitive information, privacy in meeting recordings, and potential biases in AI outputs. These findings suggest that while Copilot shows promise, its current implementation requires improvements to fully meet the diverse needs of its users.

Future research should address the limitations and gaps identified in this study. First, longitudinal studies are recommended to explore the long-term impact of Copilot on user productivity, job satisfaction, and organisational efficiency. Further investigation into training programs and user support mechanisms is critical to overcoming barriers to adoption, such as resistance to change and underutilisation of advanced features. Enhancing inclusivity, particularly in addressing multicultural contexts and reducing biases in AI-generated outputs, should also be explored and studied. Additionally, ethical frameworks need to be developed and tested to ensure secure, equitable, and transparent deployment of AI tools. Comparative studies evaluating Copilot alongside similar AI systems could provide valuable benchmarks for future improvements. Finally, user-specific adaptations and personalised AI interactions could be explored to maximise the tool's alignment with individual workflows and preferences, fostering broader acceptance and effectiveness.

# Appendix A: Pre-Trial Survey Questionnaire

From 5 choices Likert scale answer following.

1. Based on your current expectations of Copilot for Microsoft 365, please indicate your agreement or disagreement with the following statements (please scroll to the right for all options).be more efficient
2. Based on your current expectations of Copilot for Microsoft 365, please indicate your agreement or disagreement with the following statements (please scroll to the right for all options).be more productive
3. Based on your current expectations of Copilot for Microsoft 365, please indicate your agreement or disagreement with the following statements (please scroll to the right for all options).be more creative
4. Based on your current expectations of Copilot for Microsoft 365, please indicate your agreement or disagreement with the following statements (please scroll to the right for all options).save time
5. Based on your current expectations of Copilot for Microsoft 365, please indicate your agreement or disagreement with the following statements (please scroll to the right for all options).access information
6. Based on your current expectations of Copilot for Microsoft 365, please indicate your agreement or disagreement with the following statements (please scroll to the right for all options).learn new skills
7. Based on your current expectations of Copilot for Microsoft 365, please indicate your agreement or disagreement with the following statements (please scroll to the right for all options).perform better
8. Based on your current expectations of Copilot for Microsoft 365, please indicate your agreement or disagreement with the following statements (please scroll to the right for all options).stay up to date
9. Based on your current expectations of Copilot for Microsoft 365, please indicate your agreement or disagreement with the following statements (please scroll to the right for all options).generate insights
10. Based on your current expectations of Copilot for Microsoft 365, please indicate your agreement or disagreement with the following statements (please scroll to the right for all options).make informed decisions
11. Do you have any other expectations?
12. How often are you experiencing the following challenges in your current role? (please scroll to the right for all options).Excessive time spent doing one or many administrative task(s)
13. How often are you experiencing the following challenges in your current role? (please scroll to the right for all options).Challenging to find information or data
14. How often are you experiencing the following challenges in your current role? (please scroll to the right for all options).Find it hard to stay up to date (e.g. organisational updates)
15. How often are you experiencing the following challenges in your current role? (please scroll to the right for all options).Confidence writing quality emails and documents (tone, clarity, targeted messages)
16. How often are you experiencing the following challenges in your current role? (please scroll to the right for all options).Limited opportunities for innovation and creativity
17. How often are you experiencing the following challenges in your current role? (please scroll to the right for all options).Inefficiencies in getting things done (e.g. duplication of effort or rework)
18. Are you experiencing any other challenges? How frequently?
19. In your current role, how often are you using the following applications? .Outlook
20. In your current role, how often are you using the following applications? .Teams
21. In your current role, how often are you using the following applications? .Excel
22. In your current role, how often are you using the following applications? .Word
23. In your current role, how often are you using the following applications? .PowerPoint
24. Are there any other apps you are using including CSIRO applications? Please list which ones and how often.
25. While using Outlook, how much time are you spending on the following activities? (please scroll to the right for all options).Drafting emails/responding to emails
26. While using Outlook, how much time are you spending on the following activities? (please scroll to the right for all options).Reading emails (received)
27. While using Outlook, how much time are you spending on the following activities? (please scroll to the right for all options).Calendar review/management
28. While using Outlook, how much time are you spending on the following activities? (please scroll to the right for all options).Contacts review/management
29. Are there other activities that you do while using Outlook? If so, indicate what and how often.
30. While using Teams, how much time are you spending on the following activities? (please scroll to the right for all options).Real-time collaboration and communication with colleagues and/or teams
31. While using Teams, how much time are you spending on the following activities? (please scroll to the right for all options).Attending meetings
32. While using Teams, how much time are you spending on the following activities? (please scroll to the right for all options).File sharing
33. Are there other activities that you do while using Teams? If so, indicate what and how often.
34. While using Excel, how much time are you spending on the following activities? (please scroll to the right for all options).Preparing or managing data sets
35. While using Excel, how much time are you spending on the following activities? (please scroll to the right for all options).Analysing data for insights
36. While using Excel, how much time are you spending on the following activities? (please scroll to the right for all options).Summarising data
37. Are there other activities that you do while using Excel? If so, indicate what and how often.

38. While using Word, how much time are you spending on the following activities? (please scroll to the right for all options).Creating documents
39. While using Word, how much time are you spending on the following activities? (please scroll to the right for all options).Reviewing documents
40. Are there other activities that you do while using Word? If so, indicate what and how often.
41. While using PowerPoint, how much time are you spending on the following activities? (please scroll to the right for all options).Creating content/presentations
42. While using PowerPoint, how much time are you spending on the following activities? (please scroll to the right for all options).Reviewing content/presentations
43. Are there other activities that you do while using PowerPoint? If so, indicate what and how often.
44. While in your current role, how much time do you spend on the following activities? (please scroll to the right for all options).Finding information
45. While in your current role, how much time do you spend on the following activities? (please scroll to the right for all options).Staying up to date (e.g. projects, teams activities, organisational updates)
46. Are there other activities that you spend your time doing? If so, indicate what and how often. (open question)
47. Do you hold any ethical concerns in using Copilot for Microsoft 365 as part of your work? (open question)
48. Do you have any concerns about the integration of Copilot as part of your work activities and tools (word, excel, outlook etc)? (open question)

## Appendix B: Post-Trial Survey Questionnaire

From 5 choices Likert scale answer following.

1. Based on your experience using Copilot during the trial period, please indicate your agreement or disagreement with the following statements?.be more efficient
2. Based on your experience using Copilot during the trial period, please indicate your agreement or disagreement with the following statements?.be more productive
3. Based on your experience using Copilot during the trial period, please indicate your agreement or disagreement with the following statements?.be more creative and innovative
4. Based on your experience using Copilot during the trial period, please indicate your agreement or disagreement with the following statements?.save time
5. Based on your experience using Copilot during the trial period, please indicate your agreement or disagreement with the following statements?.access information that I need for my job
6. Based on your experience using Copilot during the trial period, please indicate your agreement or disagreement with the following statements?.increase my job satisfaction
7. Based on your experience using Copilot during the trial period, please indicate your agreement or disagreement with the following statements?.learn new skills faster
8. Based on your experience using Copilot during the trial period, please indicate your agreement or disagreement with the following statements?.perform better in my role
9. Based on your experience using Copilot during the trial period, please indicate your agreement or disagreement with the following statements?.stay up to date
10. Based on your experience using Copilot during the trial period, please indicate your agreement or disagreement with the following statements?.have a personalised work experience
11. Based on your experience using Copilot during the trial period, please indicate your agreement or disagreement with the following statements?.have a positive impact on my wellbeing
12. Based on your experience using Copilot during the trial period, please indicate your agreement or disagreement with the following statements?.generate insights on data
13. Based on your experience using Copilot during the trial period, please indicate your agreement or disagreement with the following statements?.make informed decisions
14. Based on your experience using Copilot during the trial period, please indicate your agreement or disagreement with the following statements?.create quality and clear communication
15. Based on your experience using Copilot during the trial period, please indicate your agreement or disagreement with the following statements?.successfully target messages
16. Based on your experience using Copilot during the trial period, please indicate your agreement or disagreement with the following statements?.read and understand information
17. Based on your experience using Copilot during the trial period, please indicate your agreement or disagreement with the following statements?.gauge other people's feelings
18. Based on your experience using Copilot during the trial period, how often are you experiencing the following challenges in your current role? .Excessive time spent doing one or many administrative ta
19. Based on your experience using Copilot during the trial period, how often are you experiencing the following challenges in your current role? .Challenging to find information or data
20. Based on your experience using Copilot during the trial period, how often are you experiencing the following challenges in your current role? .Find it hard to stay up to date (e.g. organisational updates)
21. Based on your experience using Copilot during the trial period, how often are you experiencing the following challenges in your current role? .Confidence writing quality emails and documents (tone, c
22. Based on your experience using Copilot during the trial period, how often are you experiencing the following challenges in your current role? .Limited opportunities for innovation and creativity
23. Based on your experience using Copilot during the trial period, how often are you experiencing the following challenges in your current role? .Inefficiencies in getting things done (e.g. duplication

24. What impact do you feel Copilot has had on the following activities: .Time spent doing one or many administrative task(s)
25. What impact do you feel Copilot has had on the following activities: .Finding information or data
26. What impact do you feel Copilot has had on the following activities: .Staying up to date
27. What impact do you feel Copilot has had on the following activities: .Getting things done (e.g. duplication of effort or re-work)
28. What impact do you feel Copilot has had on the following activities: .The ability to write quality emails and documents
29. What impact do you feel Copilot has had on the following activities: .Opportunities for innovation and creativity
30. Please describe any other challenges
31. While using Outlook with Copilot, how much time are you spending on the following activities? .Drafting emails/responding to emails
32. While using Outlook with Copilot, how much time are you spending on the following activities? .Reading emails (received)
33. While using Outlook with Copilot, how much time are you spending on the following activities? .Calendar review/management
34. While using Outlook with Copilot, how much time are you spending on the following activities? .Contacts review/management
35. Are there other activities that you do while using Outlook with Copilot? If so, indicate what and how often.
36. While using Teams with Copilot, how much time are you spending on the following activities? .Real-time collaboration and communication with colleagues and/or teams
37. While using Teams with Copilot, how much time are you spending on the following activities? .Attending meetings
38. While using Teams with Copilot, how much time are you spending on the following activities? .File sharing
39. Are there other activities that you do while using Teams with Copilot? If so, indicate what and how often.
40. While using Excel with Copilot, how much time are you spending on the following activities? .Preparing or managing data sets
41. While using Excel with Copilot, how much time are you spending on the following activities? .Analysing data for insights
42. While using Excel with Copilot, how much time are you spending on the following activities? .Summarising data
43. Are there other activities that you do while using Excel with Copilot? If so, indicate what and how often.
44. While using Word with Copilot, how much time are you spending on the following activities?.Creating documents
45. While using Word with Copilot, how much time are you spending on the following activities?.Reviewing documents
46. Are there other activities that you do while using Word with Copilot? If so, indicate what and how often.
47. While using PowerPoint with Copilot, how much time are you spending on the following activities? .Creating content/presentations
48. While using PowerPoint with Copilot, how much time are you spending on the following activities? .Reviewing content/presentations
49. Are there other activities that you do while using PowerPoint with Copilot? If so, indicate what and how often.
50. Using Copilot in your current role, how much time do you spend on the following activities? .Finding information
51. Using Copilot in your current role, how much time do you spend on the following activities? .Staying up to date (e.g. projects, teams activities, organisational updates, technologies and the way we do
52. Are there other activities that you spend your time doing while using Copilot? If so, indicate what and how often.
53. Did you identify any ethical concerns in using Copilot?
54. Please elaborate on your ethical concerns (open question)
55. Since using Copilot, have there been any other concerns?
56. Please elaborate on your concerns (open question)
57. What is your perception on data security and privacy while using Copilot? (open question)
58. While using Copilot, were otuputs transparent and aligned with intended references?
59. During the trial, how would you rate your experience in the following areas? .Information received prior to commencing the Copilot trial
60. During the trial, how would you rate your experience in the following areas? .The Copilot trial onboarding experience
61. During the trial, how would you rate your experience in the following areas? .Engagements while using Copilot during the trial
62. During the trial, how would you rate your experience in the following areas? .Technical support while using Copilot during the trial
63. During the trial, how would you rate your experience in the following areas? .Level of satisfaction with Copilot support and training during the trial
64. Based on what you know now about Copilot, do you think more comprehensive training and support would be required if it were to be rolled out in CSIRO? (open question)
65. Based on your experience with Copilot, would you find it valuable and beneficial to continue using in your current role? (open question)
66. How likely would you be to recommend Copilot to your colleague/s or within your professional network? (open question)
67. How do you perceive the role of AI, such as Copilot, in shaping the future of your work or industry? (open question)
68. Is there anything else you would like to share? (open question)